\documentclass[twocolumn]{aa}
\usepackage{graphicx}
\usepackage{txfonts}

\begin{document}
   \title{Time delay between images of the lensed quasar UM673}
   \titlerunning{Time delay between images of the lensed quasar UM673}
   \authorrunning{E.~Koptelova et al.}
       \author{E.~Koptelova\inst{\ref{inst1}}$^{,2,3}$\and W.
P.~Chen\inst{\ref{inst2}}\and T.~Chiueh\inst{\ref{inst1}}\and
B.P~Artamonov\inst{\ref{inst3}}\and
V.L.~Oknyanskij\inst{\ref{inst3}}\and
       S.N.~Nuritdinov\inst{\ref{inst4}}\and O.~Burkhonov\inst{\ref{inst4}}\and
T.~Akhunov\inst{\ref{inst4}}\and
       V.V.~Bruevich\inst{\ref{inst3}}\and
       O.V. Ezhkova\inst{\ref{inst3}}\and
       A.S.~Gusev\inst{\ref{inst3}}\and
       A.A.~Sergeyev\inst{\ref{inst5}}\and
Sh.A.~Ehgamberdiev\inst{\ref{inst4}}\and
M.A.~Ibragimov\inst{\ref{inst4}}
        }

   \offprints{E. Koptelova}

   \institute{Department of Physics, National Taiwan University, No.1, Sec. 4, Roosevelt Rd., 106 Taipei, Taiwan\email{ekaterina@phys.ntu.edu.tw}\label{inst1}
              \and Graduate Institute of Astronomy, Jhongli City, Taoyuan County 320, Taiwan
              \label{inst2}\and  Sternberg Astronomical Institute (SAI), Moscow State University,
              Universitetskii pr. 13, 119992 Moscow,
              Russia \label{inst3}\and Ulugh Beg Astronomical Institute of the Uzbek Academy of Sciences, Astronomicheskaya 33, 100052 Tashkent, Uzbekistan\label{inst4}
              \and Institute of Astronomy of Kharkov National University, Sumskaya 35, 61022 Kharkov, Ukraine \label{inst5}
                      }

 \date{Received February 4, 2011; accepted ...}

\abstract {}{We study brightness variations in the double lensed
quasar UM673 (Q0142-100) with the aim of measuring the time delay
between its two images.} {In the paper we combine our previously
published observational data of UM673 obtained during the 2003 --
2005 seasons at the Maidanak Observatory with archival and
recently observed Maidanak and CTIO UM673 data. We analyze the V,
R and I-band light curves of the A and B images of UM673, which
cover ten observational seasons from August 2001 to November 2010.
We also analyze the time evolution of the difference in magnitudes
between images A and B of UM673 over more than ten years.} {We
find that the quasar exhibits both short-term (with amplitude of
$\sim$0.1 mag in the R band) and high-amplitude ($\sim$0.3 mag)
long-term variability on timescales of about several months and
several years, respectively. These brightness variations are used
to constrain the time delay between the images of UM673. From
cross-correlation analysis of the A and B quasar light curves and
error analysis we measure the mean time delay and its error of
$89\pm11$ days. Given the input time delay of 88 days, the most
probable value of the delay that can be recovered from light
curves with the same statistical properties as the observed R-band
light curves of UM673 is $95^{+5}_{-16}$$^{+14}_{-29}$~days (68
and 95 \% confidence intervals). Analysis of the $V-I$ color
variations and V, R and I-band magnitude differences of the quasar
images does not show clear evidence of the microlensing variations
between 1998 and 2010.}{}


   \maketitle

\section{Introduction}

Multiple images of lensed quasars show change in their brightness
over time. There are two main reasons for these brightness
variations. One is that the quasar itself, as a variable source,
changes in brightness with time. Corresponding brightness
variations are observed in the light curves of all quasar images,
but not simultaneously. Changes in brightness in one image follow
or lead the brightness changes in the others with a certain time
lag (time delay). The time delay between these brightness
variations in the quasar images is a combination of delays that
arise due to geometrical differences between the light paths (and
thus light travel times) for each quasar image and the difference
in the gravitational potential of the lensing galaxy between image
positions. The geometrical term is related to the Hubble constant
through the angular diameter distances (see Schneider et al.
\cite{schneider}). This relation gives us a method for estimation
of the Hubble constant independently of the distance ladder
(Refsdal \cite{refsdal1964}). The potential term is determined by
the mass distribution in the lens. Thus the mass distribution of
distant galaxies can be studied using the time delays as one of
the observational constraints (see, e.g., Kochanek
\cite{kochanek2002}).

The passage of individual stars in the lensing galaxy near the
light paths of quasar images will also cause variations in
brightness known as microlensing (Chang \& Refsdal
\cite{changrefsdal1979}). These brightness variations are not
similar in each of the quasar images. The probability for
microlensing  depends on the density of stars at positions of the
images. Normally we would expect both microlensing variations and
variations intrinsic to the quasar to be present in the light
curves of the quasar images. Accurate measurement of the time
delay between the images ensures that variations due to
microlensing can be separated from the variations intrinsic to the
quasar (see Paraficz et al. \cite{paraficz2006}). However, time
delay measurement itself is often not a simple and straightforward
task. Successful measurement of the delay requires a combination
of several conditions, such as a change in the brightness of the
quasar during observations, good sampling and observational time
spans, and minimal contamination of the quasar's intrinsic
variations by variations due to microlensing.

In this study we analyze brightness variations in images of the
lensed system UM673 (Q0142-100) discovered by MacAlpine \& Feldman
(\cite{MacAlpine1982}). The system consists of a distant quasar at
redshift $z_\mathrm{q}$ = 2.719 (Surdej et al. \cite{surdej1987,
surdej1988}) gravitationally lensed by an elliptical galaxy at
redshift $z_\mathrm{l}$ = 0.49 (Surdej et al. \cite{surdej1988};
Smette et al. \cite{smette1992}, Eigenbrod et al.
\cite{eigenbrod2007}) into A and B images with the image
separation of 2\farcs2.

UM673 has been extensively observed since its discovery (Daulie et
al. \cite{daulie1993}; Sinachopoulos et al.
\cite{sinachopoulos2001}). However, earlier studies measured only
relative or integral photometry of the two of UM673 images, so the
detected brightness variations could be explained by both
intrinsic quasar variability and microlensing.

The first V and i-band light curves for each of the A and B images
of UM673 were presented in Nakos et al. (\cite{nakos2005}). The
observations showed both quasar components to be variable on
timescales ranging from several months to years. During the first
season of observations, Nakos et al. (\cite{nakos2005}) detected a
short-term event lasting for 120 days in both A and B images. It
had an amplitude of about 0.08~mag in the V band. The overall
brightness changes detected in one year of observations were 0.14
and 0.08~mag in the V and Gunn i bands, respectively. Nakos et al.
(\cite{nakos2005}) did not measure the time delay between the
images. After shifting the light curve of image B relative to that
of image A they found that the observed brightness changes of
these images did not match each other. Thus, they concluded that
either the time delay between the images was longer than 150 days
(150 days was the duration between two consecutive V-band
observations of UM673 in the 1999 and 2001 seasons) or the
brightness variations were contaminated by microlensing. Their
analysis of the $V-i$ color indices of images A and B showed that
the part of the variations in the brighter A image might be
connected to microlensing by the stars in the lensing galaxy. It
was found that image A became bluer as its brightness increased as
expected during microlensing (Wambsganss \& Paczinski
\cite{wambpacz1991}). The brightness and color variations in image
B were puzzling and could not be interpreted unambiguously.
Analysis of the UM673A\&B emission line-to-continuum ratios from
September 2002 showed them to be the same in both images, as it
would be expected in the absence of microlensing (Wisotski et al.
\cite{wisotzki2004}).

We have been conducting monitoring observations of several
gravitationally lensed systems with the aim of measuring the
lensing time delays, and to study microlensing variability (see
Koptelova et al. \cite{koptelova2005,
koptelova2006,koptelova2007}; Ull$\rm \acute{a}$n et al.
\cite{ullan2006}; Goicoechea et al. \cite{goicoechea2006,
goicoechea2008}; Shalyapin et al. \cite{shalyapin2008,
shalyapin2009}). UM673 is one of our targets. In our first paper
(Koptelova et al. \cite{koptelova2010}; Paper I) we presented an
analysis of observations of the UM673 system obtained with the
1.5-m telescope of the Maidanak Observatory (see also Koptelova et
al. \cite{koptelova2008}). Observations were made in the V, R and
I bands in the 2003, 2004 and 2005 observational seasons. The two
UM673 components brightened during the first season of
observations and then gradually faded until the end of 2005. We
interpreted the similar photometric behavior (brightening and
fading) of the A and B images as due to variability intrinsic to
the quasar. Given this assumption, the cross-correlation analysis
led to the time delay between images A and B images of about 150
days (image A is leading).

Unfortunately, the data presented in Paper I did not allow for a
detailed interpretation of the observed brightness variations. In
the current work we present new photometry and time delay analysis
of the longer observational records collected between August 2001
and November 2010. The details of the monitoring program and the
observational data are presented in Sects.~\ref{observations} and
\ref{light_curves}, respectively. Based on new observations and
analysis of the observed brightness variations we measure a
revised time delay between images A and B. The analysis of the
brightness variations in the system UM673 and the time delay
measurements are presented in Sects.~\ref{delay} and
\ref{microlensing}. A discussion is given in
Sect.~\ref{discussion}.

\section{Observations} \label{observations}

In the study we use monitoring observations of UM673 obtained
during different observational seasons at two sites.  The majority
of the observational data were collected during a quasar
monitoring program carried out by the Maidanak GLQ collaboration
(see Dudinov et al.~\cite{dudinov2000}). Images in the Bessel V, R
and I bands were obtained with the 1.5-m AZT-22 telescope of the
Maidanak Observatory (Central Asia, Uzbekistan)  during the 1998
-- 2010 observational seasons. A considerable part of these
observations, the 2003 -- 2005 data, have been presented in
Koptelova et al. (\cite{koptelova2008}) and Paper I. The V, R and
I-band observations of the lensed system were also made between
July 28, 2008 and January 18, 2010 using the 1.3-m SMARTS
telescope at CTIO, Chile (as a part of the ToO observations
carried out by National Central University, Taiwan). UM673 was
usually observed from August until December, or sometimes January,
when it was well visible at both sites. A summary of the
observational data acquired between 1998 and 2010 are given in
Table~\ref{table1}.
\begin{table*}
\caption{Summary of the UM673 observational data collected during
the 1998 -- 2010 seasons.} \label{table1} \centering
\begin{tabular}{l c c c c}
\hline    &            &                    &         & Number of
\\
Telescope & CCD camera & Bands / Exposures & Period & nights  \\
\hline

1.5-m AZT-22 &  Pitt CCD   &    V (240 s), R(240 s), I(240 s) &  Nov 1998 &  9  \\
             &  Pictor-416 CCD &    V (180 s), R(180 s), I(180 s) &  Dec 1998 &  5  \\
             &  ST-7     &    R (180 s)           &  Sep 1999 &  3  \\
             &  SITe-005 &    V (210 s), R (180 s), I (150 s) & Aug 2001 -- Jan 2006 &  134\\
             &  SNUCAM   &    V (200 s), R (200 s), I (200 s)                &
             Aug 2006 -- Nov 2010 & 92 \\
1.3-m SMARTS  &  ANDICAM  &    V(200 s), R (200 s), I (200 s)  &
Aug 2008 -- Jan 2010 & 30 \\ \hline
\end{tabular}
\end{table*}

The Maidanak data were obtained with different CCD cameras
installed at the 1.5-m telescope. During the 1998 observational
season images were obtained with the TI 800 x 800 Pitt and
Pictor-416 CCD cameras with pixel scales of 0.13 and 0.16 arcsec
pixel$^{-1}$, respectively. The 1999 images were obtained with the
ST-7 760 x 510 pixel CCD provided by the Maidanak Foundation (see
Dudinov et al.~\cite{dudinov2000}). The field of view (FOV) of the
images taken with these three CCD cameras was small so images did
not include any bright stars in the vicinity of UM673, which are
useful for performing differential photometry of the UM673 A and B
quasar images. Between August 2001 and August 2006 images were
obtained with the 2000 x 800 pixel SITe-005 CCD camera
manufactured in the laboratory of Copenhagen University. The
images taken in long-focus and short-focus modes have pixel scales
of 0.135 and 0.268 arcsec pixel$^{-1}$, respectively. The most
recent observational data were obtained with a new 4096~x~4096
SNUCAM camera provided by Seoul University. The images taken with
this CCD camera have pixel scale of 0.266 arcsec pixel$^{-1}$ and
FOV of $18\farcm1$ x $18\farcm1$. The characteristics and
performance of SNUCAM on the 1.5-m telescope are discussed in
detail in Im et al.~\cite{im2010}. The 1.3-m SMARTS telescope
obtained images using the dual-channel optical/near-infrared CCD
camera ANDICAM which has an optical FOV of 6\farcm3 x 6\farcm3
(0.369 arcsec pixel$^{-1}$). On each observational night images
were taken in a series of 2 -- 8 frames in all three V, R and I
bands.

\section{UM673 A and B light curves} \label{light_curves}
The V, R and I-band photometry of UM673 from August 2001 to
November 2010 is now discussed. In the current work we revisit
photometry of UM673 between the 2003 -- 2005 observational seasons
presented in Koptelova et al. (\cite{koptelova2008}) and Paper I
(magnitudes of the A and B images of UM673 for this period are
given in Table 2 of Koptelova et al. \cite{koptelova2008}), and
perform photometry of the 2001 and 2006 -- 2010 data. The
photometry method we use is the PSF fitting method and it has been
described in our first paper. In the current analysis we improve
the accuracy of photometry in the following ways. First, we find
that performance of the PSF fitting method is poor when applied to
the individual frames of UM673. The fainter components of UM673
has low signal-to-noise ratio, especially in the new 2006 -- 2010
data when the quasar was faint. For example, the signal to noise
ratio of the A and B images of UM673 in the 2006 R-band data is
estimated to be about 200 and 70, respectively. The low
performance results in high level of statistical and correlated
errors. We find that these errors are more severe for the 2006 --
2010 data and significantly affect the cross-correlation analysis
of the 2006 -- 2010 light curves producing spurious peaks at short
time lags. Therefore, in order to minimize the errors, we apply
the PSF fitting method to the combined frames of UM673. The
combined frames are a sum of two or three individual frames with
similar seeing taken at the same observational night. Usually we
sum up three sequential frames from the same night series. For
small fraction of nights, when the system was observed only two
times, we sum up two frames of UM673. This allows us to enhance
the signal-to-noise ratio of the images by a factor of $\sqrt{3}$
or $\sqrt{2}$. Second, in the current analysis we choose several
isolated stars around UM673 to construct the PSF model. The shape
of the PSF can vary over image plane as a result of optical
aberrations in the telescope and camera system. The aberrations
will distort the PSF shape from the center outwards. We estimate
that this effect is more severe for the new 2006 -- 2010 Maidanak
data taken with the large area 4096 x 4096 pixel SNUCAM CCD. UM673
is usually not in the center of the SNUCAM frames. In this case,
the PSF constructed from several nearby stars is a better
representation of the PSF shape at the location of UM673 on the
image frame. The PSF was constructed using the bright star
north-east of UM673, labeled as star 1 (see Fig. 1 in Paper I),
and stars 1 and 3 from the catalog of secondary standard stars
around lensed quasars (see Nakos et al. \cite{nakos2003}). In this
way, the photometric analysis was conducted in the same manner for
the whole data set. The measured fluxes were calibrated relative
to star 1 introduced in Paper I. The magnitudes of star 1 in the
V, R and I bands are $m_\mathrm{V} = 14.653 \pm 0.008$,
$m_\mathrm{R} = 14.278 \pm 0.008$ and $m_\mathrm{I} = 13.954 \pm
0.009$ mag, respectively.

There might be a contribution of the lensing galaxy into the flux
of the closest image B. It is estimated to be negligible in the V
band ($m_\mathrm{gal}^{\rm{V}} = 20.81$ mag (Leh$\rm \acute{a}$r
et al. \cite{lehar2000})). The galaxy contribution in the R and
I-band fluxes of image B are measured to be 0.069 and 0.126 mag,
respectively (see Koptelova et al. \cite{koptelova2008}).

\begin{figure*}
\centering
\includegraphics[width=15cm]{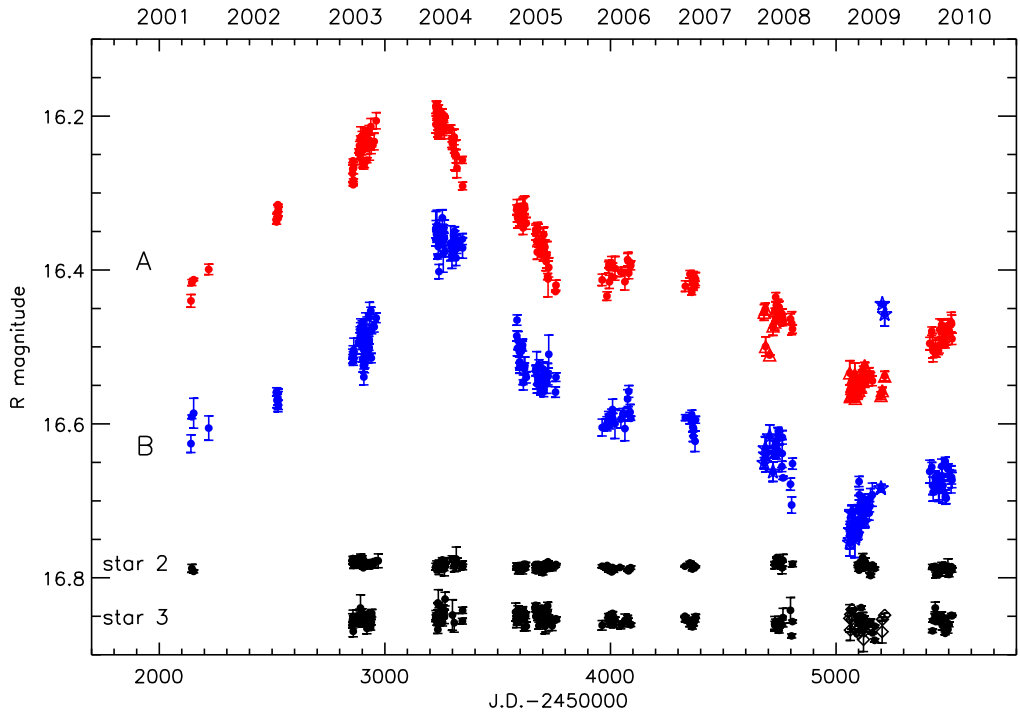}
\caption {R-band light curves of the A and B images of UM673 from
August 2001 to November 2010. For better representation, the light
curve of image B is shifted by -1.87 mag. The light curves of
reference stars 2 and 3 are shown at the bottom.} \label{fig1_col}
\end{figure*}

\begin{figure*}
\centering
\includegraphics[width=15cm]{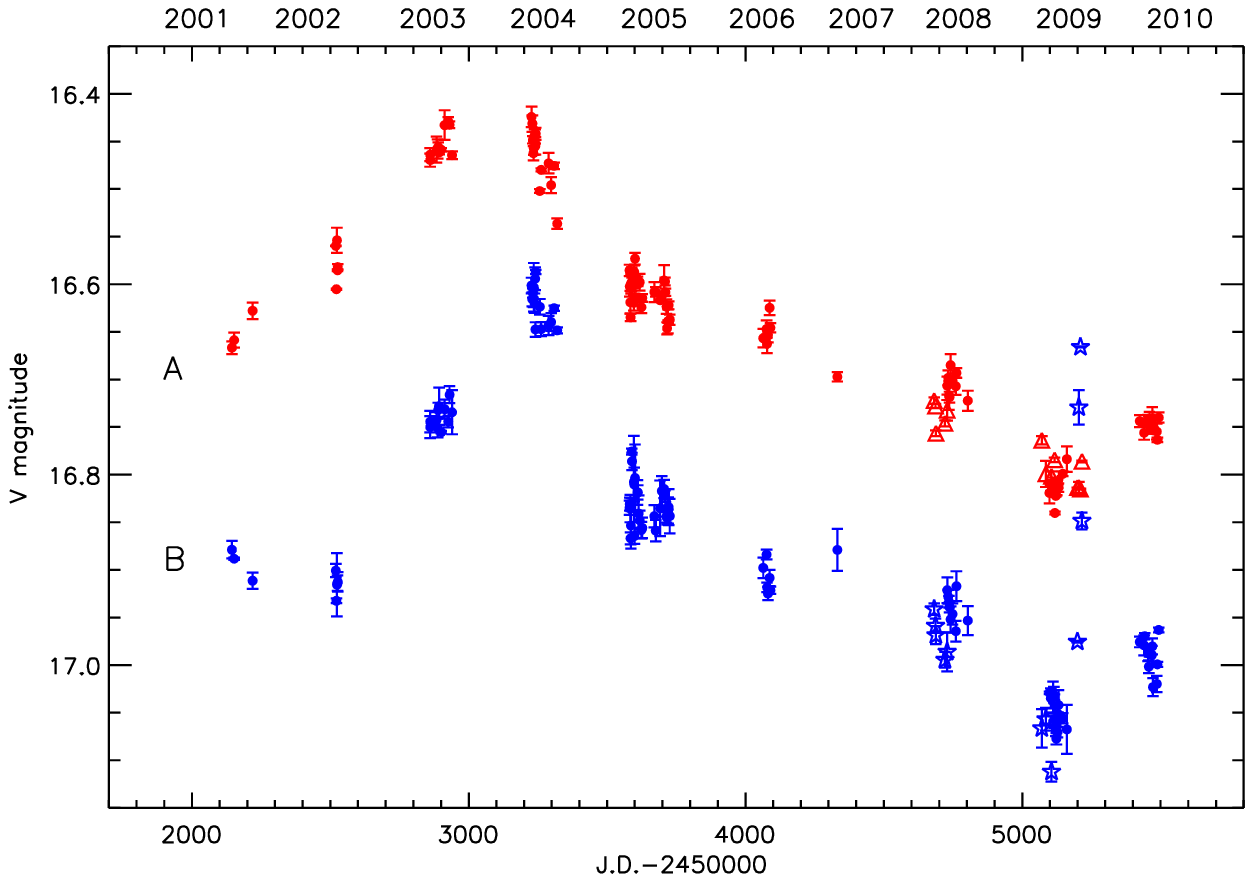}
\caption {V-band light curves of the A and B images of UM673 from
August 2001 to November 2010. The light curve of image B is
shifted by -1.95 mag.} \label{fig2_col}
\end{figure*}

\begin{figure*}
\centering
\includegraphics[width=15cm]{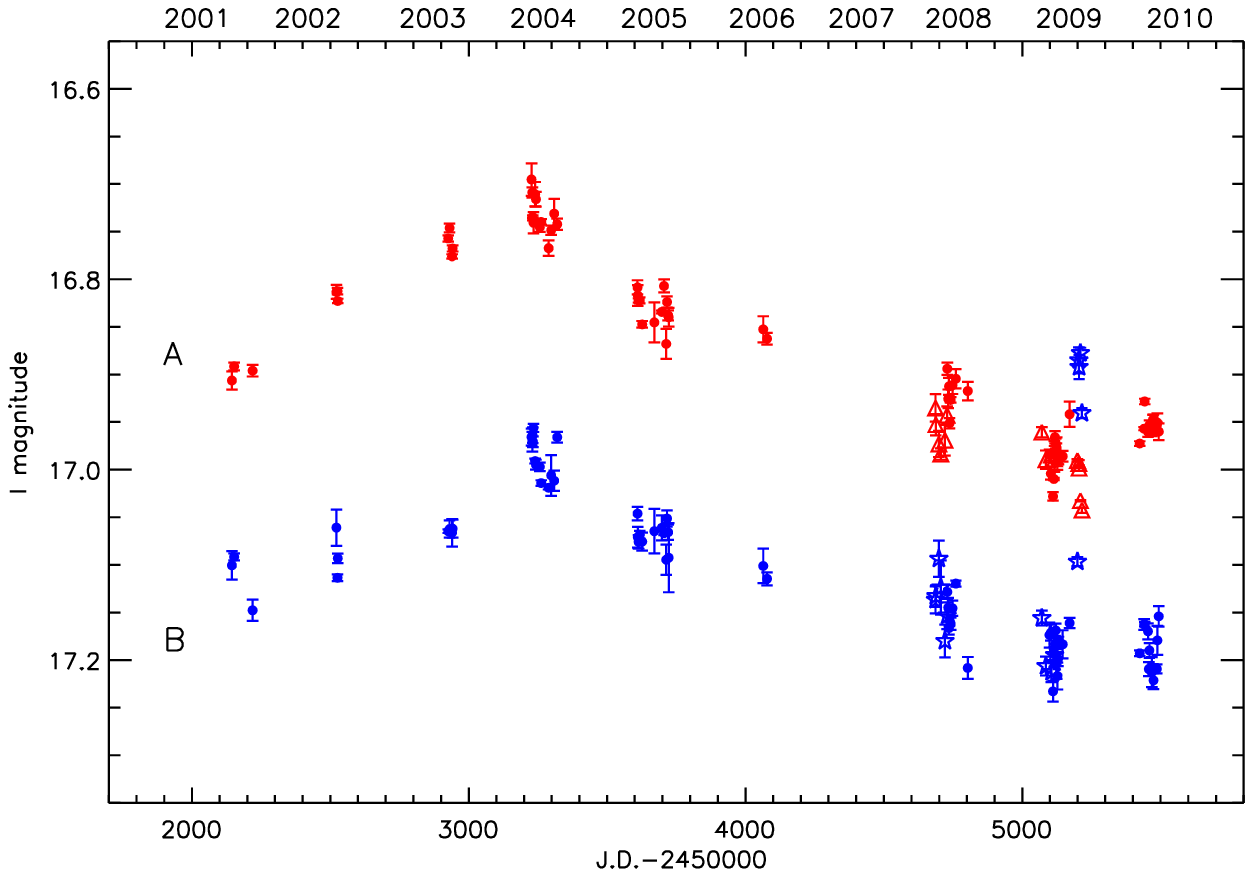}
\caption {I-band light curves of the A and B images of UM673 from
August 2001 to November 2010. The light curve of image B is
shifted by -1.62 mag.} \label{fig3_col}
\end{figure*}

The resulting Maidanak and CTIO R-band light curves of the A and B
quasar images are shown in Fig.~\ref{fig1_col}. The filled circles
indicate the Maidanak data points for images A and B,
respectively. The triangles and stars indicate the CTIO data
points obtained during the 2008 -- 2010 seasons for images A and
B, respectively. The V and I-band light curves presented in
Figs.~\ref{fig2_col} and ~\ref{fig3_col} are available in the
electronic version of the paper. Transmission properties of the
CTIO filters are slightly different from the Bessel filters used
in the Maidanak observations\footnote{Transmission curves of the
CTIO filters in comparison with the Bessel filters can be found at
http://www.ctio.noao.edu/telescopes/50/1-3m.html}. We measure that
magnitude differences between star 3 and bright star 1 in the
Maidanak images are 2.850, 2.878 and 2.892~mag in the V, R and I
bands, respectively. The corresponding differences in the CTIO
images are 2.916, 2.942 and 2.977~mag in the V, R and I bands,
respectively. The CTIO light curves are matched to the Maidanak
light curves taking into account these differences in the relative
magnitudes of stars 1 and 3 in the different bands. The
photometric errors of the individual measurements were estimated
as standard deviations of the mean values of several measurements
made on each observational night. The mean standard errors of the
photometry in the R band are estimated to be
$\overline{\sigma}_\mathrm{A} = 0.007$ and
$\overline{\sigma}_\mathrm{B} = 0.010$~mag for images A and B,
respectively. Fig.~\ref{fig1_col} also shows the light curves for
reference stars 2 and 3 in the field of view of UM673. Stars are
labeled as in our first paper (see Fig. 1 in Paper I). The fluxes
of these two stars were measured relative to calibration star 1.
From our data, the R-band magnitude of star 2 is $m_\mathrm{R} =
16.246 \pm 0.008$ mag. Photometry of other nearby stars in the
field of UM673, including star 3 ($m_\mathrm{R} = 17.196 \pm
0.014$), was presented in Nakos et al. (\cite{nakos2003}). The
field of view of the images taken with the 1.3-m SMARTS telescope
is smaller and does not include star 2 or some other bright stars
seen in the Maidanak images. Therefore, for the CTIO images we
plot the relative photometry of star 3 and bright star 1
(indicated by the open rhomboids in Fig.~\ref{fig1_col}).

As can be seen in Fig.~\ref{fig1_col}, the global brightness
changes of both components of UM673 are similar over the course of
our observations. Along with the long-term brightness changes
which take years, we can detect short-term brightness variations
on timescales of several months. These variations on different
timescales probably have a different origin. The global long-term
behavior of the light curves with the largest brightness changes
(more than 0.3 mag in the R band) might be connected with the
formation and evolution of the accretion disk (e.g., Lyuty
\cite{lyuty2006}, Ar\'{e}valo et al. \cite{arevalo2009}). In these
long-term brightness variations we can distinguish a global
maximum and minimum observed in both light curves in 2004 and in
2009, respectively. The short-term brightness variations might be
due to reprocessing of the X-ray flares by the accretion disk (see
Krolik et al. \cite{krolik1991}). Short-term brightness variations
of the UM673 images on timescales of several months have been
previously detected by Nakos et al. (\cite{nakos2005}). Between
January 2 and 18, 2010 we detect fast high-amplitude variability
event in image B of UM673. The amplitude of this event is several
times higher than it is usually observed on similar timescales
within the observational seasons. If this brightness variation is
a result of quasar variability then similar variation has to be
seen first in image A. However, we do not detect similar event in
the image A light curves during the same observational season.
Although there is no other variation like the image B event in the
A and B light curves, it might not be unique for the lensed
system. Analysis of the flux ratios between image A and B in 1998
shows another evidence of short-term variations of comparable
amplitude. This event is discussed later in
Sect.~\ref{microlensing}. In the next section, we perform
cross-correlation analysis of the A and B light curves to measure
the time delay between the quasar images.

%

%
%
%
%
%

\section{Intrinsic quasar variability and time delay analysis} \label{delay}

Earlier estimate of the time delay between the A and B images of
UM673 was made based on slow long-term brightness changes observed
between the 2003 -- 2005 seasons. Analysis of the better-sampled
R-band light curves gave a time delay and conservative error of
$150^{+7}_{-18}$$^{+42}_{-36}$~days (confidence levels of 68 and
95~\%) (see Paper I). The estimated delay of 150 days is
comparable to one season of observations of the lensed system at
the Maidanak Observatory. The light curves of the A and B quasar
images shifted by this delay did not overlap each other. This made
it difficult to verify the obtained result.

The observations of UM673 also show noticeable brightness changes
within each observational season. These short-term brightness
variations observed in the quasar images have not been considered
carefully in Paper I given the assumption that the quasar
brightness does not change significantly on short timescales.
Moreover, there are more features in the global behavior of the
light curves than it was available before. The quasar seems to
reach the minimum of its brightness in 2009 and starts gradually
brightening again. In our current time delay analysis we consider
these long-term high-amplitude brightness variations showing a
maximum in 2004 and a minimum in 2009. In the analysis we also
take into account the short-term variations in brightness within
each observational season.

The time delay is measured with the modified cross-correlation
function (MCCF) method (see Oknyanskij \cite{oknyanskij1993}). The
method, its application and test performance for analysis of time
series containing large annual gaps are described in Paper I.
Here, we briefly outline the approach. In the MCCF method, each
data point from the B light curve, $B(t_{\rm i})$, forms a pair
with an interpolated point from the A light curve, $A(t_{\rm
i}+\tau)$ at time $t_{\rm i}+\tau$, where $\tau$ is the time lag.
The pairs of data points for which $\tau-\Delta t_{\rm} \leq
\Delta t_{\rm ij} < \tau+\Delta t_{\rm }$ (where $\Delta t_{\rm
ij} = \mid t_{\rm j}-t_{\rm i}\mid$ is the time shift between the
$t_{\rm i}$ point of the $A$ light curve and the $t_{\rm j}$ point
of the $B$ light curve) are then used to calculate the
cross-correlation function. The interpolation interval $\Delta
t_{\rm}$ is usually chosen as a compromise between the desire to
decrease the interpolation errors and to find a sufficient number
of data pairs to reliably calculate the correlation coefficient
for a given time lag.

For the analysis of the light curves presented in Paper I the
value of $\Delta t_{\rm }$ was adopted to be 90 days under the
assumption that quasar UM673 is a slow variable source. This was
the smallest value of $\Delta t_{\rm }$ that we could choose given
the large annual gaps in the light curves. The timescale of the
short-term variations is comparable, or sometimes, shorter than
the interpolation interval of 90~days. In this case the MCCF
method does not allow for taking into account variations which are
shorter than 90 days, it becomes less sensitive to the short-term
variations. In addition, interpolation errors produced for large
values of $\Delta t_{\rm }$ can lead to an erroneous time delay
estimate.

In this work, in order to account for the short-term brightness
changes and minimize the errors, we use two interpolation
intervals, $\Delta t_{\rm max}$ and $\Delta t_{\rm min}$. The
interpolation interval $\Delta t_{\rm max}$ = 90 days is the same
interval adopted for calculations of the CCF in Paper I. The
interpolation interval $\Delta t_{\rm min}$ is introduced to take
into account the short-term quasar variations within each
observational seasons. It is used to calculate the
cross-correlation function for those data pairs for which both
data points in the pair (the real point from the B light curve and
the interpolated one from the A light curve) are within the same
observational season. When the data points do not lie within the
same season of observations, $\Delta t_{\rm max}$ is used instead
of $\Delta t_{\rm min}$. This approach is applied to calculate the
cross-correlation function between the time-shifted interpolated A
light curve and the discrete B light curve. The time lag $\tau$
ranges from $-500$ to 500 days with a step of 1 day. A value of 10
days chosen for $\Delta t_{\rm min}$ is comparable to the average
sampling of the light curves within one observational season. The
origin of high-amplitude rapid brightness changes observed in
image B in January, 2010 is unclear. It can be either intrinsic to
the quasar with the counterpart in image A which was not observed,
or unique for image B. To avoid influence of the data points
corresponding to this event on correlation between the A and B
light curves, they were excluded from the time delay analysis.

\begin{figure}
\resizebox{\hsize}{!}{\includegraphics{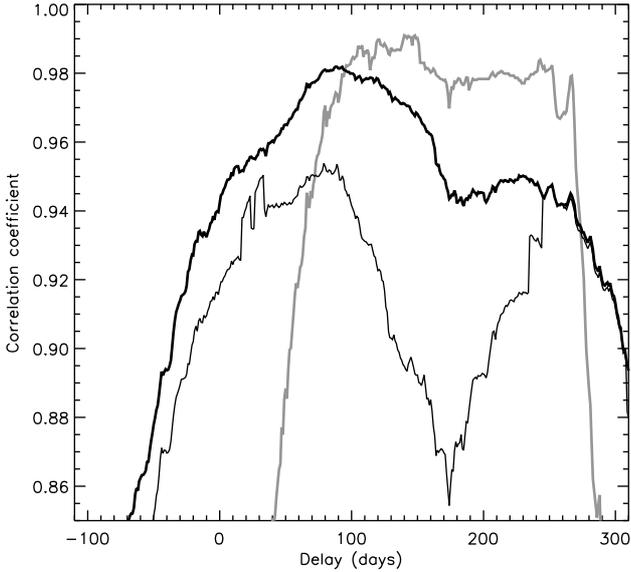}} \caption{CCFs
calculated between the R-band light curves of images A and B with
the data points corresponding to the high-amplitude event in image
B excluded (shown by a thick black line); and the data points
corresponding to the high-amplitude event in image B included
(shown by a thin black line). The CCF calculated between the
R-band light curves corresponding to the 2003 -- 2005 period is
shown by a thick grey line.} \label{fig4}
\end{figure}
\begin{figure}
\resizebox{\hsize}{!}{\includegraphics{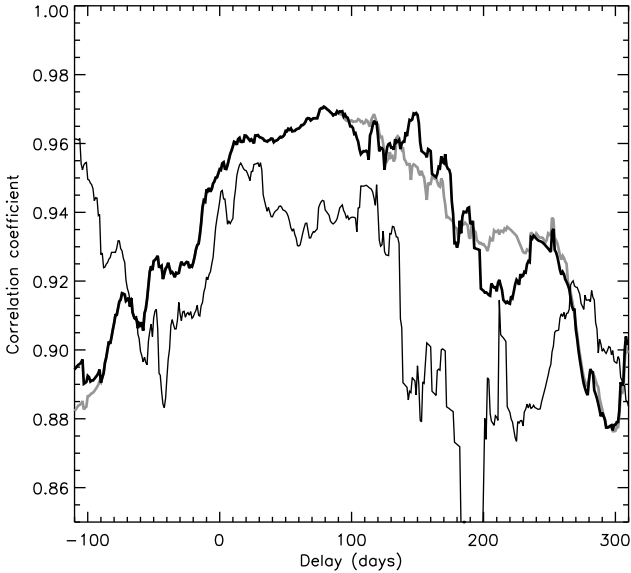}} \caption{CCFs
calculated between the V (thick black line) and I-band (thin black
line) light curves of the A and B images. Thick grey line shows
the V-band CCF calculated for the value of $\Delta t_{\rm max} =
110$ days.} \label{fig5}
\end{figure}

The resulting CCFs for the R, V and I-band data are shown in
Figs.~\ref{fig4} and \ref{fig5}. The CCF calculated between the
better-sampled R-band light curves (shown by a thick black line in
Fig.~\ref{fig4}) reaches its maximum at a delay of 88 days with a
correlation coefficient of 0.981. For the comparison we plot the
CCF calculated between the R-band light curves with the data
points corresponding to the high-amplitude event in image B
included (shown by a thin black line in Fig.~\ref{fig4}). As can
be seen, this CCF also reaches its maximum at a delay of about 88
days but with lower correlation coefficients. It is not as smooth
as the first CCF and has secondary peaks at short delays. We
interpret these short-delay peaks as originating from the
high-amplitude event in image B. The grey thick line in
Fig.~\ref{fig4} shows the CCF calculated between segments of the
R-band light curves which include only the data points collected
between 2003 and 2005. The same observational data have been used
to measure the time delay between the images of UM673 in Paper I.
The corresponding CCF has different shape and reaches its maximum
at a very different delay of 142 days. The most probable reason
for the disagreement in the results is that the single
parabola-shape long-term brightness variation observed between
2003 and 2005 can not reliably constrain the delay. The broad peak
of this CCF falls in the range of time delays for which the A and
B light curves do not overlap each other. We find that in order to
measure the delay which is longer or comparable to observational
seasons, it is important to analyze more features in the global
behavior of the light curves, rather than one single event.
Altogether, the long-term brightness changes provide a better
constrain on the time delay.

\begin{figure*}
\centering
\includegraphics[width=15cm]{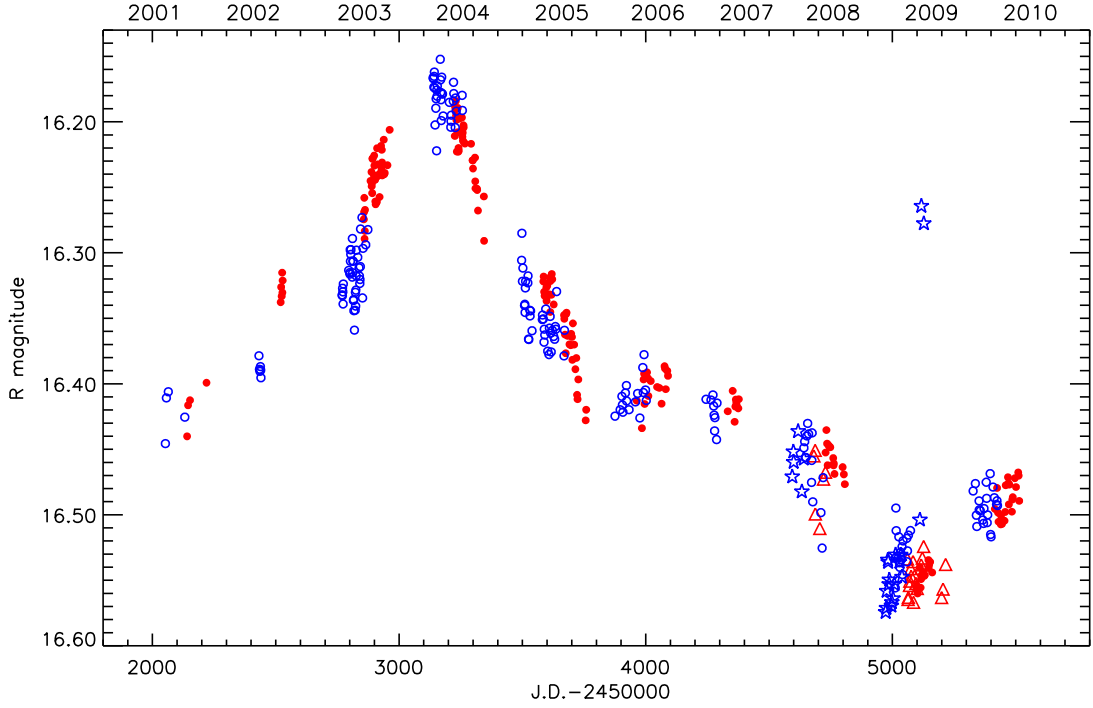}
\caption {R-band light curves of image A (filled circles) and
image B shifted by a time delay of 88 days with a magnitude offset
of -2.12~mag (open circles).} \label{fig7_col}
\end{figure*}

As can be seen in Fig.~\ref{fig5}, the shapes of the V and I-band
CCFs are different from the shape of the R-band CCF (especially
for the I-band light curves). The CCF calculated between the
V-band light curves reaches its maximum at a delay of 79 days. The
V-band CCF has several secondary peaks on the top of the main
peak. The secondary peaks at longer delays disappear with the
increase of the interpolation interval. This is demonstrated with
the CCF calculated for the value of $\Delta t_{\rm max} = 110$
days (shown by a thick grey line in Fig.~\ref{fig5}). Therefore,
we conclude that these secondary peaks are most probably artifacts
caused by the errors in calculation of the CCF at longer time
lags. For these lags the number of data pair contributing to the
calculation of CCF is small, therefore the accuracy of the CCF is
low. The other features of the V-band CCF corresponding to the
more prominent central peak remain unchanged. The I-band CCF
reaches its maximum at a very different time lag of about 20~days,
although it also has secondary peaks at longer delays. Apparently,
the poorly sampled I-band light curves with smaller amplitudes of
the brightness changes can not accurately constrain delays which
are longer than duration of the observational seasons in the I
band. As a result, the MCCF method can not find enough data pairs
to reliably calculate the CCF at delays longer than 20 days. This
leads to a decrease of the I-band CCF at longer delays.

From the cross-correlation analysis we find that the V and R-band
CCFs give consistent time delays, although slightly different. As
the R-band light curves are better sampled and the R-band time
delay corresponds to a higher value of the correlation coefficient
than the V-band delay, we consider the R-band value of the delay
as a more robust measurement. The light curves of the UM673 images
corrected for the time delay of 88 days and the magnitude offset
of 2.12 mag are shown in Fig.~\ref{fig7_col}. For ease of
presentation the errorbars in the A and B light curves are not
shown. We find a good match in the global behavior of the light
curves. There is also an overlap of about two months between the
light curves for most of the observational seasons.

\begin{figure}
\resizebox{\hsize}{!}{\includegraphics{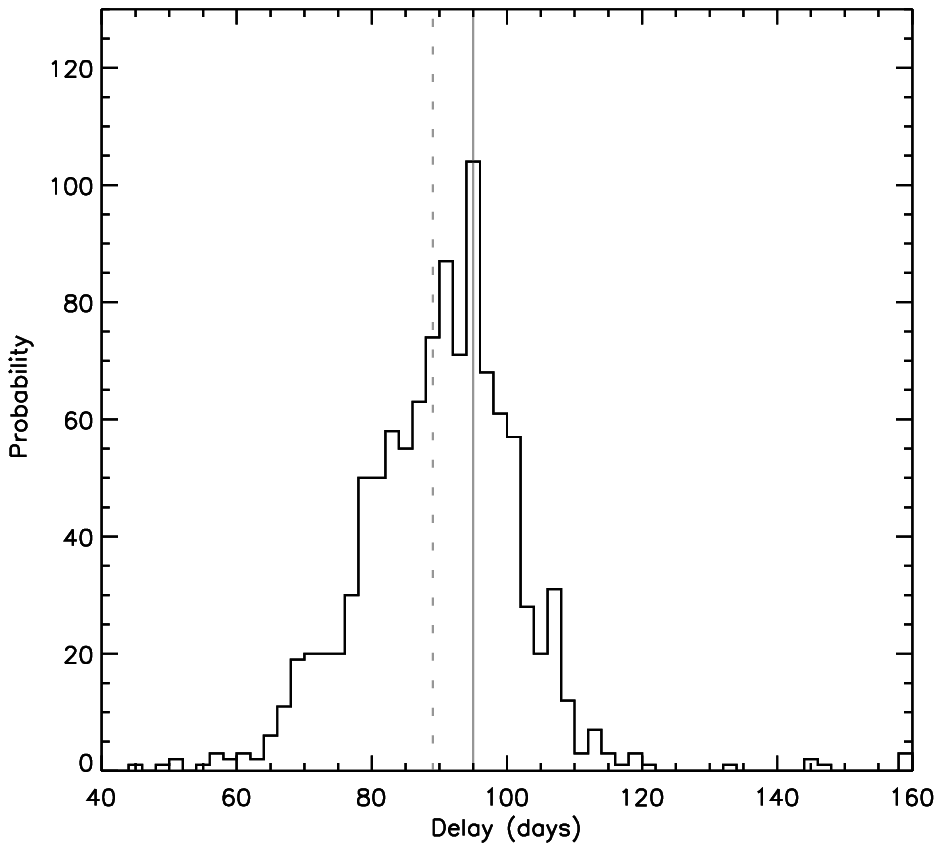}}
\caption{Distribution of peaks of the CCF obtained for 1000 Monte
Carlo realizations of the R-band light curves. The peak of the
distribution is marked by a solid grey line. The dotted line
corresponds to the mean value of the delay. } \label{fig10}
\end{figure}

Uncertainties in time delay measurement due to photometric errors
and systematic sampling effects are investigated with the Monte
Carlo simulations. We perform simulations of 1000 artificial light
curves using Timmer \& Koenig's algorithm
(\cite{timmerkoenig1995}) (these simulations are discussed in
detail in Paper I). The distribution of the time delays recovered
from cross-correlation analysis of the Monte Carlo simulated
R-band light curves of images A and B, shifted by the input time
delay of 88 days, is shown in Fig.~\ref{fig10}. From this
distribution we find the mean time delay and its error of
$89\pm11$~days (marked by a dotted line in Fig.~\ref{fig10}). The
most probable value of the delay that can be measured from light
curves  with similar statistical properties and variability
pattern as the observed R-band light curves is
$95^{+5}_{-16}$$^{+14}_{-29}$ days (68 and 95 \% confidence
intervals).

%

%
%

\begin{figure*}
\centering
\includegraphics[width=15cm]{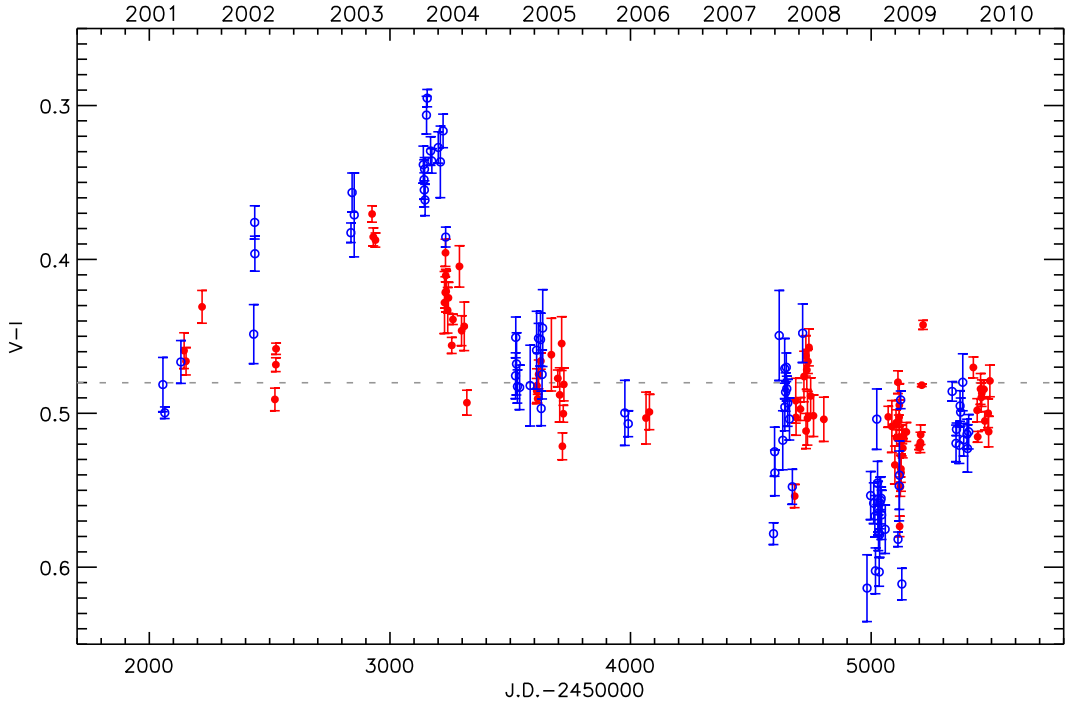}
\caption {$V-I$ color curves of images A (filled circles) and B
(open circles) of UM673. The color curve of image B is shifted by
a time delay of 88 days and a magnitude offset of $-$0.326 mag.
The dotted line traces the average color of the quasar.}
\label{fig11_col}
\end{figure*}

\section{Color variations and evolution of flux ratio} \label{microlensing}

In this section we analyze color variations and evolution of the
flux ratio of the UM673 images over more than ten years. The $V-I$
color light curves of the A and B images of UM673 between August
2001 and November 2010 are shown in Fig.~\ref{fig11_col}. The
color changes are expected to be similar in both images separated
by the time delay. In Fig.~\ref{fig11_col} the image B light curve
is shifted by the time delay of 88 days and corrected for the
$V-I$ color difference between images A and B of about 0.326 mag.
The combined light curve represents the $V-I$ color variations of
the quasar in the period from 2001 to 2010. As can be seen in
Figs.~\ref{fig7_col} and \ref{fig11_col}, the color variations are
well correlated with the brightness variations of the quasar.
Similar to the brightness changes, the color curve also shows
global maximum and minimum in 2004 and 2009, respectively. From
the brightness and color  variations of the quasar we see that the
image B light curve recorded the brightest state of the quasar.
During maximum of the brightness, the quasar was bluer than on
average and during minimum of the brightness the quasar was redder
than on average. The overall change in the $V-I $ color index is
about 0.3 mag. The overall change the brightness is more than 0.4
mag in the V band. The correlation between color and brightness
variations of the quasar is in agreement with numerous
observations which show that quasars are generally
bluer-when-brighter (see, e.g., Tr{\`{e}vese et al.
\cite{trevese2001}; Wilhite et al. \cite{wilhite2005}).

We also analyze the archive Maidanak images of UM673 taken in the
V, R and I bands in November, 1998. These image frames have small
field of view and do not contain any bright stars except the
lensed system UM673. Analysis of these data gives only the
relative magnitudes of images A and B of UM673 in the V, R and I
bands. We use these measurements of the relative fluxes to
estimate the difference between the $V-I$ color indices of the
images in 1998, calculated as $\Delta(V-I)_{\mathrm BA} = \Delta
m_{\mathrm BA}^{\mathrm V}-\Delta m_{\mathrm BA}^{\mathrm I}$. The
measured color difference $\Delta(V-I)_{\mathrm BA}$ for November,
1998 is $0.373\pm0.014$~mag. We find that it is in close agreement
with the mean color difference between the UM673 images measured
based on the 2001 - 2010 data. Therefore, the color difference
between the quasar images is the same on average over more than
ten years. This can be considered as an evidence that there were
no noticeable microlensing variations in the images of UM673.

In addition, we analyze the differences in magnitude (flux ratios)
between images A and B in the V, R and I bands at different
epochs. The relation between the magnitude difference and flux
ratio is given by $\Delta m(B-A) = 2.512*\lg(F_{\mathrm
A}/F_{\mathrm B})$. For the analysis we use the following data:
images of UM673 taken with the EMMI camera of the ESO New
Technology Telescope in 1998 \footnote{These images were acquired
during Engineering programme 1, Proposal No. 59.A-9001(A).};
images obtained with FORS1 at the ESO Very Large Telescope in
2000\footnote{PI/CoI J. Hjorth et al. Proposal No. 65.O-0666(C).},
2004\footnote{PI/CoI Meylan et al. Proposal No. 074.A-0563(A).}
and 2006\footnote{PI/CoI Meylan et al. Proposal No.
077.A-0155(B).}; archive Maidanak data collected during the 1998
and 1999 seasons and more recently, between 2001 and 2010 (see
Table~\ref{table1}). Note that the ESO data used in the analysis
were obtained in the same Bessel system of filters as the Maidanak
data. Fig.~\ref{fig12} shows the flux ratios between images A and
B over 12 years. The flux ratios measured using the ESO images for
the V, R and I-band data are marked by open stars, open triangles
and open squares, respectively. The Maidanak flux ratios for the
V, R and I-band data are indicated by open circles, filled circles
and stars, respectively. The R-band Maidanak flux ratio measured
between 2003 and 2010 is corrected for the time delay of 88 days.
The rest of the data is too sparse or taken only at a single epoch
to calculate time-delay corrected flux ratios. The Maidanak-CTIO
flux ratios which are not corrected for the time delay are
estimated as average flux ratios for each observational season.
The ESO flux ratios are measured based on single-epoch
observations. The flux ratio errors are estimated as follows. From
the A and B light curves, rms amplitudes of the quasar variability
for each season of observations are in the range of $0.010\leq
\sigma_{\mathrm var}^{\mathrm R}\leq0.031$ mag in the R band. In
order to account for possible changes in the quasar brightness on
timescales of 88 days we add $\sigma_{\mathrm var}^{\mathrm R}
\simeq 0.031$ in quadrature to weighted average errors measured
for each of the observational seasons (see also Shalyapin et al.
\cite{shalyapin2009})
\begin{figure}
\resizebox{\hsize}{!}{\includegraphics{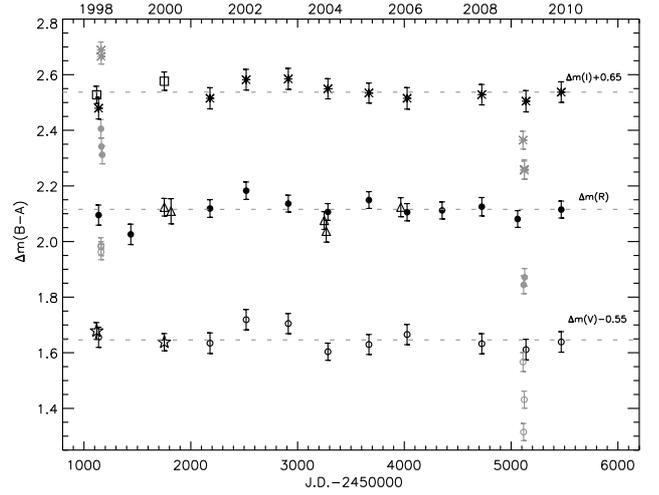}}
\caption{Differences in the V, R and I-band magnitudes between
images A and B of UM673 for the 1998 -- 2010 period. The
measurements based on the ESO archive data are indicated by open
stars, open triangles and open squares for the V, R and I bands,
respectively. The Maidanak-CTIO magnitude differences are shown by
open circles, filled circles and stars. The mean V, R and I-band
magnitude differences are shown by dotted lines.} \label{fig12}
\end{figure}

We estimate that the flux ratios were $\Delta m^{\mathrm V}$=2.19,
$\Delta m^{\mathrm R}$=2.11 and $\Delta m^{\mathrm I}$=2.03 mag
for the ESO VLT observations in July, 2000. They roughly agree
with the Maidanak V, R and I-band flux ratios of the quasar images
in 1998 and 1999. The magnitude differences for the earlier
HST/WFPC2 data obtained in 1994 (see Keeton et al.
\cite{keeton1998}) are $\Delta m^{\mathrm F555W}$=2.24, $\Delta
m^{\mathrm F675W}$=2.29 and $\Delta m^{\mathrm F814W}$=2.09~mag
(where the F555W, F675W and F814W HST/WFPC2 bands roughly match
the standard Johnson-Cousins V, R and I bands). Note that although
the F555W, F675W and F814W HST/WFPC2 filters are relatively good
approximations of the standard V, R and I filters, the difference
in photometry between these two photometric systems can reach 0.1
mag (Holtzman et al. \cite{holtzman1995}). The Maidanak V, R,
I-band flux ratios measured based on the multiepoch data collected
between 1998 and 2010 are $\Delta m^{\mathrm V}=2.19\pm0.04$,
$\Delta m^{\mathrm R}=2.12\pm0.04$ and $\Delta m^{\mathrm
I}=1.98\pm0.04$ mag. The I-band flux ratio of 1.98~mag is in a
good agreement with the NIR K and $\rm{L'}$-band flux ratios
measured by Fadely \& Keeton (\cite{fadely2011}) based on single
epoch observations (1.92 and 1.89~mag, respectively).

As can be seen in Fig.~\ref{fig12}, the V, R and I-band flux
ratios were stable at different epochs. This can be seen better
from the R-band flux ratio measurements corrected for the time
delay. The stability of the flux ratio in the different bands
indicates the absence of microlensing-induced variations in the
system. Small deviations of the flux ratio, which is not corrected
for the time delay, from its multiepoch mean value can be
explained by the variations intrinsic to the quasar.

It seems that the quasar can exhibit high-amplitude variations
over short timescales. From analysis of the archive Maidanak data
we find evidence of rapid brightness variations in the system
between November 13 and December 23, 1998. The V, R, and I-band
relative fluxes of the A and B images changed significantly over a
very short time. In particular, the weighted average magnitude
differences of images between November 13 and November 26, 1998
were $\Delta m^{\mathrm V}=2.21\pm0.04$, $\Delta m^{\mathrm
R}=2.07\pm0.04$ and $\Delta m^{\mathrm I}=1.99\pm0.04$~mag.
However, in the next month (between December 8 and December 23,
1998) the weighted average magnitude differences were already
$\Delta m^{\mathrm V}=2.53\pm0.04$, $\Delta m^{\mathrm
R}=2.27\pm0.04$ and $\Delta m^{\mathrm I}=2.05\pm0.04$~mag. These
measurements are shown by grey symbols in Fig.~\ref{fig12}. The
increase in magnitude difference $\Delta m(B-A)$ (of about 0.3 mag
in the V band) might indicate significant brightening of image A
or simultaneous fading of image B and brightening of image A in
December, 1998. The R-band magnitude difference measured from data
obtained in September 1999, showed that it returned to the value
of November 1998: $\Delta m^{\mathrm R}=2.03\pm0.04$~mag. The
magnitude difference calculated for this event shows a clear
dependence on the wavelength. The change of the V-band flux ratio
is more prominent than that of the I-band flux ratio. Nakos et al.
(\cite{nakos2005}) published photometric results for nearly the
same epoch of observations. During the time interval covered by
the observations, the A and B light curves of UM673 showed the
presence of rapid short-term variations in both images (see Figs.
3 and 4 in Nakos et al. \cite{nakos2005}). These rapid variations
could alter single epoch flux ratio during a very short time. The
same brightness variations should be observed in image B after the
88 day time delay. However this cannot be confirmed due to the
lack of observational data for this period.

The high-amplitude brightness variation observed in image B in
January, 2010 does not have its counterpart in image A. Taking
into account the time delay of 88 days, it should be seen in the A
light curve in the beginning of October, 2009. As we do not
observe the same brightness variation in image A, it is most
probably not connected with the intrinsic quasar variations. The
amplitude of brightness changes during this event is higher in the
V band (about 0.39 mag) than in the I band (about 0.31 mag) as
expected for the microlensing variations (see Wambsganss \&
Paczynski \cite{wambpacz1991}). However, microlensing by the stars
in the lensing galaxy would take much longer time. We conclude
that an independent confirmation of this event might be needed to
find an explanation for its origin.

\section{Discussion} \label{discussion}

In this study we present the V, R and I-band light curves of the A
and B images of the lensed quasar UM673. The light curves cover
ten observational seasons, from August, 2001 to November, 2010. We
find that both images of UM673 show brightness variations on short
(several months) and long (several years) timescales in all three
bands. Using cross-correlation analysis of the better-sampled
R-band light curves we estimate the mean time delay between images
A and B (image A is leading) and its error to be $89\pm11$~days.
From the Monte Carlo simulations, the most probable value of the
delay that can be measured from light curves with similar
statistical properties and variability pattern as the observed
R-band light curves is $95^{+5}_{-16}$$^{+14}_{-29}$ days (68 and
95 \% confidence intervals). These measurements are based on the
observations of much longer time coverage than in Paper I. The
time delay of about 150 days measured in Paper I was constrained
based on the long-term parabola-shape brightness variation
observed between 2003 and 2005. We find that this single event
does not allow for correct determination of the time delay. For
the revised time delay of 89 days, the global behavior of the A
and B light curves matches well. This demonstrates that the
observed brightness variations are mainly due to intrinsic
variations of the quasar. Analysis of the brightness and color
changes does not show evidence of the microlensing variations. The
bluer-when-brighter behavior of image A found in the earlier
observations of Nakos et al. (\cite{nakos2005}) is most probably
due to the quasar variability rather than due to microlensing.

We find that the flux ratio between the quasar images corrected
for the time delay does not evolve with time. Therefore, it is not
altered by microlensing which would otherwise causes changes in
the flux ratio with time. The measured mean flux ratios
$F_{\mathrm A}/F_{\mathrm B}$ are 7.6, 7.1 and 6.3 in the V, R and
I bands, respectively. The estimated V-band mean flux ratio is in
good agreement with the value of Wisotzki et al.
(\cite{wisotzki2004}). In Wisotzki et al. (\cite{wisotzki2004})
the spectrum of image B was rescaled by a factor of 7.78 to match
the C IV emission line of image A. Therefore, the estimated
emission-line flux ratio between the images is found to be 7.78 at
5780 $\AA$, which roughly corresponds to the effective wavelengths
of the V filter. Since there is no microlensing, the difference in
the flux ratio in the V, R and I bands is most probably due to
extinction in the lensing galaxy (Yonehara et al.
\cite{yonehara2008}).

The measured time delay can be used to estimate the Hubble
parameter and constrain the mass model of the lensing galaxy.
There are several lens models which predict different time delays
between the UM673 images.  The predicted time delay for the lens
with elliptical symmetry and $H_{0}=75$~km~s$^{-1}$Mpc$^{-1}$ is
about 7 weeks (Surdej et al. \cite{surdej1988}). Leh$\rm
\acute{a}$r et al. (\cite{lehar2000}) fitted a set of four
standard lens models (SIE, constant M/L models, and those with
external shear). The SIE and constant M/L models predict time
delay $h \Delta t$ = 80 and $h \Delta t$ = 121~days, respectively.
The SIE and constant M/L models with external shear predict time
delay $h \Delta t$ = 84 -- 87 and $h \Delta t$ = 115~days,
respectively. Given that $\Delta t=89$~days, the Hubble constant
$\rm{H_{0}^{meas}}$ estimated for the SIE and M/L models is 90 and
136 $\rm{~km~s^{-1} Mpc^{-1}}$, respectively. For the SIE and M/L
models with shear it is 94 and 129 $\rm{~km~s^{-1} Mpc^{-1}}$,
respectively. These values of the Hubble constant are higher than
the Hubble key project result of $72 \pm 8 \rm{~km~s^{-1}
Mpc^{-1}}$ (Freedman et al. \cite{freedman2001}) or improved
result of $74.2 \pm 3.6 \rm{~km~s^{-1} Mpc^{-1}}$ (Riess et al.
\cite{riess2009}). This might be a result of an additional
convergence to the lensing potential from nearby objects or
objects on the line of sight to the quasar (see, e.g., Keeton et
al. \cite{keeton2000}). If we take into account the total external
convergence $k_{\rm{T}}$ of the nearby objects observed in the
field of view of UM673, the Hubble parameter from the SIS model,
corrected as $\rm{H_{0} = (1-k_{T})H_{0}^{meas}}$, is $78 \pm
10~\rm{km s^{-1} Mpc^{-1}}$. This value roughly agrees within the
errors with the Hubble key project value of the Hubble parameter.
However, there still might be an unaccounted convergence produced
by objects on the line of sight to the quasar.

Recently, Cooke et al. (\cite{cooke2010}) reported the discovery
of a previously unrecognized DLA system at z=1.63 in the spectrum
of image A of UM673. They also found a weak Ly$\alpha$ emission
line in the spectrum of image B at the same redshift as the DLA
that indicates a star formation rate of 0.2 solar mass per year.
The discovery provides evidence of an additional mass, a galaxy
which gives rise to the DLA system toward the UM673 quasar.

The accuracy of the Hubble constant from the time delay in UM673
can be improved in the future by analyzing the external
convergence produced by the objects in the field of view of UM673
and reducing the error in the time delay measurement. The latter
requires coordinated observations of UM673 at different sites over
time interval which can provide better overlap between time delay
corrected light curves of the quasar images than the Maidanak-CTIO
data. UM673 might exhibit rapid brightness variations of more than
0.1 mag on timescales from one to several months. Observations of
these rapid brightness variations during coordinated monitoring of
the system can help to reduce uncertainty in the time delay down
to several per cent.

\begin{acknowledgements}
We would like to thank Vyacheslav Shalyapin for helpful
discussions. This research was supported by the Taiwan National
Science Councils grant No. NSC99-2811-M-002-051. We also
gratefully acknowledge the support of the Russian Foundation for
Basic Research (RFBR, grant No. 09-02-00244a) for travel of the
SAI team to the Maidanak Observatory. The research was also
supported by a grant for young scientists from the President of
the Russian Federation (No. MK-2637.2006.2), and the Deutscher
Akademischer Austausch Dienst (DAAD) grant No. A/05/56557.
\end{acknowledgements}


\begin{thebibliography}{}



\bibitem[2009]{arevalo2009} Ar\'{e}valo,~P., Uttley,~P., Lira,~P., et al. 2009, \mnras, 397,
2004


\bibitem[1979]{changrefsdal1979} Chang,~K., \& Refsdal,~S. 1979, \nat, 282, 561


\bibitem[2010]{cooke2010} Cooke, R., Pettini,~M., Steidel, C.C., King,~L.J., Rudie, G.C.,  \& Rakic, O. 2010, \mnras, 409,
679


\bibitem[1993]{daulie1993} Daulie,~G., Hainaut,~O., Hutsem$\rm \acute{e}$kers,~D., et al. 1993, Gravitational Lenses in the Universe, in Proc. 31st Liege International
Astrophysical Colloquium, ed. J.~Surdej, D. Fraipont-Caro, E.
Gosset, S.~Refsdal, \& M.~Remy (Universite de Liege, Institut
d'Astrophysique, Liege), 181

\bibitem[2000]{dudinov2000} Dudinov, V., Bliokh, P., Paczynski, et al. 2000, Kin.\&Phys.Cel.Bodies, 3,
170


\bibitem[2007]{eigenbrod2007} Eigenbrod,~A., Courbin,~F.,\& Meylan,~G. 2007, \aap, 465,
51



\bibitem[2011]{fadely2011} Fadely,~R., \& Keeton,~C.R. 2011, \aj, 141,
101



\bibitem[2001]{freedman2001} Freedman,~W.L., Madore,~B.F., Gibson,~B.K., et al. 2001, \apj, 553,
47


\bibitem[2006]{goicoechea2006} Goicoechea,~L. J., Ull$\rm \grave{a}$n,~A., Ovaldsen, J. E., et al.
2006, in Highlights of Spanish Astrophysics IV, ed. F.~Figueras,
J.M.~Girart, M.~Hernanz \& C.~Jordi (Springer, Dordrecht), CD-ROM
(astro-ph/0609647)


\bibitem[2008]{goicoechea2008} Goicoechea,~L.J., Shalyapin,~V.N., Koptelova,~E., et al.  2008, \na,
13, 182


\bibitem[1995]{holtzman1995} Holtzman,~J.A., Burrows,~C.J., Casertano,~S., et al. 1995, \pasp,
107, 1065



\bibitem[2010]{im2010}  Im,~M., Ko,~J., Cho,~Y., et al. 2010, JKAS,
43, 75


\bibitem[1991]{krolik1991} Krolik,~J.H., Horne,~K., Kallman,~T.R., et al. 1991, \apj, 371, 541


\bibitem[1998]{keeton1998} Keeton,~C.R., Kochanek,~C.S., \& Falco,~E.E. 1998, \apj, 509,
561

\bibitem[2000]{keeton2000} Keeton,~C.R., Christlein,~D., Zabludoff, A.I. 2000, \apj, 545,
129


\bibitem[2002]{kochanek2002} Kochanek,~C.S. 2002, \apj, 578,
25


\bibitem[2005]{koptelova2005} Koptelova,~E., Shimanovskaya,~E., \& Artamonov,~B. 2005, \mnras, 356, 323

\bibitem[2006]{koptelova2006} Koptelova,~E. A., Oknyanskij,~V.L., \& Shimanovskaya,~E. V. 2006,
\aap, 452, 37


\bibitem[2007]{koptelova2007} Koptelova,~E.A., Artamonov,~B.P., Shimanovskaya,~E.V., et al. 2007,
Astronomy Reports, 51, 797



\bibitem[2008]{koptelova2008} Koptelova,~E., Artamonov,~B.P., Bruevich,~V.V., Burkhonov,~O. A., Sergeev,~A.V., 2008, Astronomy Reports, 52, 270


\bibitem[2010]{koptelova2010} Koptelova,~E., Oknyanskij,~V.L., Artamonov~B.P., Burkhonov~O. 2010, \mnras,
401, 2805 (Paper I)

\bibitem[2000]{lehar2000} Leh$\rm \acute{a}$r,~J., Falco,~E.E., Kochanek,~C.S., et al. 2000, \apj, 536, 584

\bibitem[2006]{lyuty2006} Lyuty,~V.M.  2006, in AGN Variability from X-Rays to Radio Waves ASP Conference
Series, Vol. 360, ed. C.M.~Gaskell, I.M.~McHardy, B.M.~Peterson \&
S.G.~Sergeev (Astronomical Society of the Pacific, San Francisco),
3

\bibitem[1982]{MacAlpine1982} MacAlpine,~G. M., Feldman,~F.R. 1982, \apj, 261, 412




\bibitem[2003]{nakos2003} Nakos, Th., Ofek,~E.O., Boumis,~P.,~Cuypers,~J., Sinachopoulos,~D., van Dessel,~E., Gal-Yam,~A., Papamastorakis,~J. 2003, \aap, 402,
1157


\bibitem[2005]{nakos2005} Nakos, Th., Courbin, F., Poels, J., et al. 2005, \aap, 441,
443

\bibitem[1993]{oknyanskij1993} Oknyanskij, V. L. 1993, Pis'ma Astron. Zh., 19,
   1021

\bibitem[2006]{paraficz2006} Paraficz,~D., Hjorth,~J., Burud,~I., Jakobsson,~P., \& El$\rm \acute{i}$asd$\rm \acute{o}$ttir,~$\rm \acute{A}$. 2006,
  \aap, 455, L1

\bibitem[1964]{refsdal1964} Refsdal, S. 1964, \mnras, 128, 307


\bibitem[2009]{riess2009} Riess,~A.G., Macri,~L., Casertano,~S., et al. 2009, \apj, 699,
539


\bibitem[1992]{schneider} Schneider, P., Ehlers,~J., \& Falco, E.E. 1992, Gravitational Lenses,
(Springer, Berlin)


\bibitem[2001]{sinachopoulos2001} Sinachopoulos,~D., Nakos,~Th., Boumis,~P., et al. 2001, \apj, 122, 1692




\bibitem[2008]{shalyapin2008} Shalyapin,~V.N., Goicoechea,~L.J., Koptelova,~E., Ull$\rm \grave{a}$n, A., \& Gil-Merino,~R. 2008, \aap,
492, 401


\bibitem[2009]{shalyapin2009} Shalyapin,~V.N., Goicoechea,~L.J., Koptelova,~E., et al. 2009, \mnras,
397, 1982


\bibitem[1992]{smette1992} Smette,~A., Surdej,~J., Shaver,~P.A., et
al. 1992, \apj, 389, 39

\bibitem[1987]{surdej1987} Surdej, J., Magain, Swings,~J.-P., et al.
1987, Nature, 329, 695

\bibitem[1988]{surdej1988} Surdej, J., Magain, Swings,~J.-P., et al.
1988, \aap, 198, 49


\bibitem[1995]{timmerkoenig1995} Timmer,~J., \& K\"{o}nig,~M. 1995, \aap, 300, 707



\bibitem[2001]{trevese2001} Tr\`{e}vese,~D., Kron,~R.G, Bunone,~A., 2001, \apj, 551, 103





\bibitem[2006]{ullan2006} Ull$\rm \acute{a}$n,~A., Goicoechea,~L.J., Zheleznyak,~A.P., et al.  2006, \aap,
452, 25


\bibitem[1991]{wambpacz1991} Wambsganss,~J., \& Paczinski~B. 1991, \apj, 102, 864



\bibitem[2005]{wilhite2005} Wilhite,~B.C., Vanden Berk,~D.E., Kron,~R.G., et. al. 2005,
\apj, 633, 638


\bibitem[2004]{wisotzki2004} Wisotzki,~L., Becker,~T., Christensen,~L., et al. 2004, Astron. Nachr., 325, 135




\bibitem[2008]{yonehara2008} Yonehara,~A., Hirashita,~H., \& Richter,~P.
2008, \aap, 478, 95




\end{thebibliography}
\end{document}